\documentclass[12pt,thmsa]{article}
\usepackage{sw20aip}

\input tcilatex

\begin{document}

\title{Path integral for a pair of time-dependent coupled and driven oscillators}
\author{F. Benamira and L. Guechi \\
Laboratoire de Physique Th\'{e}orique,\\
D\'{e}partement de Physique,\\
Facult\'{e} des Sciences, Universit\'{e} Mentouri,\\
Constantine 25000 DZ, Algeria.}
\date{\today}
\maketitle

\begin{abstract}
The propagator for a certain class of two time-dependent coupled and driven
harmonic oscillators with time-varying angular frequencies and masses is
evaluated by path integration. This is simply done through suitably chosen
generalized canonical transformations and without presupposing the
knownledge of any auxiliary equation. The time-dependent oscillators system
with an exponentially growing masses and coupling coefficient in time may be
considered as particular case.

PACS 03.65. Ca - Formalim

PACS 03.65. Db - Functional analytical method
\end{abstract}

${{}^2}$%
The Feynman path integral formalism is known to be an elegant and powerful
approach for analytically treating some time-dependent one dimensional
physical systems. We can quote , for instance, the harmonic oscillator with
variable frequency or mass and a perturbative force \cite
{Khan,Gerry,Dhara,Khan2,Bin} , the infinite potential well with moving wall 
\cite{Chetouani}, the time-dependent potentials\cite{Duru} of particular
type $V(x-f(t))$ and the time-dependent model \cite{Grosche} of the form $%
V(q/\varsigma (t))/\varsigma ^{2}(t)$ with $\varsigma
(t)=(at^{2}+bt+c)^{1/2} $. We wish to add to this list the system of two
time-dependent coupled and driven harmonic oscillators. No exact solution of
this problem by path integration is known to us. On the other hand, there
has been two attempts to solve this kind of problems. The first one has been
devoted to a system composed by two time-dependent coupled oscillators with
damping and time-dependent external force\cite{Bose} . The resolution of the
Schr\"{o}dinger equation has been made by means of the construction of a
time-dependent invariant which is linear in position and momentum variables,
but the solution has not proved satisfactory since, the coupling coefficient
does not appear in the wave function. The other one relative to a system of
two time-dependent coupled and driven harmonic oscillators with time-varying
angular frequencies and masses\cite{Lo} . The propagator has been built
through a variant of the $su(1,1)$ Lie algebraic approach, but it is only
valid in the case where the masses are constant and starting with a
rotational transformation parametrized by a constant angle.

Our intention in this paper is to calculate the propagator for this system
within the framework of the path integral approach. The calculation is
facilitated by utilizing the Hamiltonian formalism and the generalized
canonical transformations.

For the dynamical system of our interest (see Ref.\cite{Lo}), the
Hamiltonian has the form: 
\begin{equation}
H(t)=\stackunder{j=1}{\stackrel{2}{\sum }}\left[ \frac{p_{j}^{2}}{2m_{j}(t)}+%
\frac{1}{2}m_{j}(t)\omega _{j}^{2}(t)x_{j}^{2}-m_{j}(t)f_{j}(t)x_{j}\right]
+\lambda (t)x_{1}x_{2},  \label{a.1}
\end{equation}
where $m_{j}(t)$, $\omega _{j}(t)$, $f_{j}(t)$ and $\lambda (t)$ are the
time-dependent mass, angular frequency, force and coupling, respectively.
This Hamiltonian is a generalization of the Hamiltonian system considered in
Ref.\cite{Bose} , where $m_{j}(t)=m_{j}e^{\gamma t}$ and $\lambda
(t)=rm_{1}m_{2}e^{\gamma t}.$

The quantum mechanical evolution of the system can be described by the
propagator, in the phase space formulation of Feynman's path integral, which
is defined formally by 
\begin{equation}
K(x_1^{\prime \prime },x_2^{\prime \prime },t^{\prime \prime };x_1^{\prime
},x_2^{\prime },t^{\prime })=\int \mathcal{D}x_1\mathcal{D}x_2\mathcal{D}p_1%
\mathcal{D}p_2\exp \left\{ \frac i\hbar S\right\} ,  \label{a.2}
\end{equation}
where 
\begin{equation}
S=\int_{t^{^{\prime }}}^{t^{^{\prime \prime }}}\left( p_1\stackrel{.}{x}%
_1+p_2\stackrel{.}{x}_2-H(t)\right) dt  \label{a.3}
\end{equation}
is the Hamilton's principal action integral.

In a time-graded representation, this expression (\ref{a.2}) is understood
as: 
\begin{eqnarray}
K(x_{1}^{\prime \prime },x_{2}^{\prime \prime },t^{\prime \prime
};x_{1}^{\prime },x_{2}^{\prime },t^{\prime }) &=&\stackunder{N\rightarrow
\infty }{\lim }\int \stackunder{n=1}{\stackrel{N-1}{\prod }}dx_{1n}dx_{2n}%
\stackunder{n=1}{\stackrel{N}{\prod }}\frac{dp_{1n}}{2\pi \hbar }\frac{%
dp_{2n}}{2\pi \hbar }  \nonumber \\
&&\times \exp \left\{ \frac{i}{\hbar }\stackunder{n=1}{\stackrel{N}{\sum }}%
S_{n}\right\} ,  \label{a.4}
\end{eqnarray}
with a short-time action 
\begin{equation}
S_{n}=\left( p_{1n}\triangle x_{1n}+p_{2n}\triangle x_{2n}-\varepsilon H(%
\widetilde{t}_{n})\right) .  \label{a.5}
\end{equation}
Here, we have adopted the standard notation: $\varepsilon
=t_{n}-t_{n-1}=\left( t^{\prime \prime }-t^{\prime }\right) /N,$ $\widetilde{%
t}_{n}=\left( t_{n}+t_{n-1}\right) /2,\quad x_{jn}=x_{j}(t_{n}),\quad
x_{j}^{\prime }=x_{j}(t^{\prime })=x_{j}(t_{0}),$ $\quad x_{j}^{\prime
\prime }=x_{j}(t^{\prime \prime })=x_{j}(t_{N});\quad j=1,2.$

The path integral (\ref{a.2}) is not trivial but can be evaluated. In order
to cast the integral into a more tractable form, it is advantageous to use
the method of canonical transformations by performing the time-dependent
canonical transformation $(x_{1},x_{2},p_{1},p_{2},t)\rightarrow
(Q_{1},Q_{2},P_{1},P_{2},t)$ defined as: 
\begin{equation}
\left\{ 
\begin{array}{c}
x_{1}=\frac{Q_{1}\cos \alpha (t)+Q_{2}\sin \alpha (t)}{\sqrt{m_{1}(t)}}%
,\qquad x_{2}=\frac{-Q_{1}\sin \alpha (t)+Q_{2}\cos \alpha (t)}{\sqrt{%
m_{2}(t)}}, \\ 
p_{1}=\sqrt{m_{1}(t)}\left( P_{1}\cos \alpha (t)+P_{2}\sin \alpha (t)+\beta
_{1}(t)x_{1}\right) , \\ 
p_{2}=\sqrt{m_{2}(t)}\left( -P_{1}\sin \alpha (t)+P_{2}\cos \alpha (t)+\beta
_{2}(t)x_{2}\right) ,
\end{array}
\right.  \label{a.6}
\end{equation}
where the functions $\alpha (t),$ $\beta _{1}(t)$ and $\beta _{2}(t)$ can be
conveniently chosen to make separation of variables straightforward possible.

From the classical mechanics equations \cite{Goldstein} 
\begin{equation}
p_{j}=\frac{\partial }{\partial x_{j}}F_{2}(x_{1},x_{2},P_{1},P_{2},t),\quad
Q_{j}=\frac{\partial }{\partial P_{j}}F_{2}(x_{1},x_{2},P_{1},P_{2},t),\text{
}j=1,2,  \label{a.7a}
\end{equation}
\qquad and 
\begin{equation}
\mathcal{H}(t)=H(t)+\frac{\partial }{\partial t}F_{2},  \label{a.7b}
\end{equation}
the generating function responsible for the transformation is found to be 
\begin{eqnarray}
F_{2}(x_{1},x_{2},P_{1},P_{2},t) &=&\sqrt{m_{1}(t)}\left( P_{1}\cos \alpha
(t)+P_{2}\sin \alpha (t)\right) x_{1}  \nonumber \\
&&+\sqrt{m_{2}(t)}\left( -P_{1}\sin \alpha (t)+P_{2}\cos \alpha (t)\right)
x_{2}  \nonumber \\
&&+\frac{1}{2}\left[ \beta _{1}(t)\sqrt{m_{1}(t)}x_{1}^{2}+\beta _{2}(t)%
\sqrt{m_{2}(t)}x_{2}^{2}\right] .  \label{a.8}
\end{eqnarray}

In the new conjugate variables $(P_{1},P_{2},Q_{1},Q_{2}),$ the Hamiltonian
for the system becomes

\begin{eqnarray}
\mathcal{H}(t) &=&\frac{1}{2}\left( P_{1}^{2}+P_{2}^{2}\right)
+A(t)P_{1}Q_{1}+B(t)P_{2}Q_{2}+C(t)\left( P_{1}Q_{2}+P_{2}Q_{1}\right)  
\nonumber \\
&&\ +\stackrel{.}{\alpha }(t)\left( P_{1}Q_{2}-P_{2}Q_{1}\right) +\frac{1}{2}%
D_{1}(t)Q_{1}^{2}+\frac{1}{2}D_{2}(t)Q_{2}^{2}+E(t)Q_{1}Q_{2}  \nonumber \\
&&-F_{1}(t)Q_{1}-F_{2}(t)Q_{2},  \label{a.9}
\end{eqnarray}
where the time-dependent coefficients $A(t),$ $B(t),$ $C(t),$ $D_{1}(t),$ $%
D_{2}(t),$ $E(t),$ $F_{1}(t)$ and $F_{2}(t)$ are given as 
\begin{equation}
\left\{ 
\begin{array}{c}
A(t)=\left( \frac{\beta _{1}(t)}{\sqrt{m_{1}(t)}}+\frac{\stackrel{.}{m_{1}}%
(t)}{2m_{1}(t)}\right) \cos ^{2}\alpha (t)+\left( \frac{\beta _{2}(t)}{\sqrt{%
m_{2}(t)}}+\frac{\stackrel{.}{m_{2}}(t)}{2m_{2}(t)}\right) \sin ^{2}\alpha
(t), \\ 
B(t)=\left( \frac{\beta _{1}(t)}{\sqrt{m_{1}(t)}}+\frac{\stackrel{.}{m_{1}}%
(t)}{2m_{1}(t)}\right) \sin ^{2}\alpha (t)+\left( \frac{\beta _{2}(t)}{\sqrt{%
m_{2}(t)}}+\frac{\stackrel{.}{m_{2}}(t)}{2m_{2}(t)}\right) \cos ^{2}\alpha
(t), \\ 
C(t)=\left[ \frac{1}{2}\left( \frac{\stackrel{.}{m_{1}}(t)}{m_{1}(t)}+\frac{%
\stackrel{.}{m_{2}}(t)}{m_{2}(t)}\right) +\left( \frac{\beta _{1}(t)}{\sqrt{%
m_{1}(t)}}-\frac{\beta _{2}(t)}{\sqrt{m_{2}(t)}}\right) \right] \sin \alpha
(t)\cos \alpha (t) \\ 
D_{1}(t)=d_{1}(t)\cos ^{2}\alpha (t)+d_{2}(t)\sin ^{2}\alpha (t)-\frac{%
2\lambda (t)\sin \alpha (t)\cos \alpha (t)}{\sqrt{m_{1}(t)m_{2}(t)}}, \\ 
D_{2}(t)=d_{1}(t)\sin ^{2}\alpha (t)+d_{2}(t)\cos ^{2}\alpha (t)+\frac{%
2\lambda (t)\sin \alpha (t)\cos \alpha (t)}{\sqrt{m_{1}(t)m_{2}(t)}}, \\ 
E(t)=\left( d_{1}(t)-d_{2}(t)\right) \sin \alpha (t)\cos \alpha (t)+\frac{%
\lambda (t)\left( \cos ^{2}\alpha (t)-\sin ^{2}\alpha (t)\right) }{\sqrt{%
m_{1}(t)m_{2}(t)}}, \\ 
F_{1}(t)=\sqrt{m_{1}(t)}f_{1}(t)\cos \alpha (t)-\sqrt{m_{2}(t)}f_{2}(t)\sin
\alpha (t), \\ 
F_{2}(t)=\sqrt{m_{1}(t)}f_{1}(t)\sin \alpha (t)+\sqrt{m_{2}(t)}f_{2}(t)\cos
\alpha (t),
\end{array}
\right.   \label{a.10}
\end{equation}
with 
\begin{equation}
d_{1}(t)=\omega _{1}^{2}(t)+\frac{\beta _{1}^{2}(t)}{m_{1}(t)}+\frac{1}{%
m_{1}(t)}\frac{d}{dt}\left( \sqrt{m_{1}(t)}\beta _{1}(t)\right) \text{,}
\label{a.11}
\end{equation}
and

\begin{equation}
d_2(t)=\omega _2^2(t)+\frac{\beta _2^2(t)}{m_2(t)}+\frac 1{m_2(t)}\frac
d{dt}\left( \sqrt{m_2(t)}\beta _2(t)\right) .  \label{a.12}
\end{equation}
As a result of these transformations, the action (\ref{a.2}) takes the form:

\begin{eqnarray}
S &=&\int_{t^{\prime }}^{t^{\prime \prime }}\left( -Q_{1}\stackrel{.}{P}%
_{1}-Q_{2}\stackrel{.}{P}_{2}-\mathcal{H}(t)+\frac{dF_{2}}{dt}\right) dt 
\nonumber \\
&=&\frac{1}{2}\left[ \left. \beta _{1}(t)\sqrt{m_{1}(t)}x_{1}^{2}+\beta
_{2}(t)\sqrt{m_{2}(t)}x_{2}^{2}\right| _{t^{\prime }}^{t^{\prime \prime }}%
\right]  \nonumber \\
&&+\int_{t^{\prime }}^{t^{\prime \prime }}\left( P_{1}\stackrel{.}{Q}%
_{1}+P_{2}\stackrel{.}{Q}_{2}-\mathcal{H}(t)\right) dt.  \label{a.13}
\end{eqnarray}
By means of a procedure similar to that presented in Ref. \cite{Chetouani} ,
the measure of the path integral in (\ref{a.2}) changes as

\begin{equation}
\mathcal{D}x_{1}\mathcal{D}x_{2}\mathcal{D}p_{1}\mathcal{D}p_{2}=\stackunder{%
j=1}{\stackrel{2}{\prod }}\left[ m_{j}^{\prime \prime }m_{j}^{\prime }\right]
^{\frac{1}{4}}\mathcal{D}Q_{j}\mathcal{D}P_{j},  \label{a.14}
\end{equation}
where $m_{j}^{\prime \prime }=m_{j}(t^{\prime \prime })$ and $m_{j}^{\prime
}=m_{j}(t^{\prime }).$ It therefore follows from (\ref{a.13}) and (\ref{a.14}%
) that the propagator (\ref{a.2}) may be written in the form: 
\begin{eqnarray}
K(x_{1}^{\prime \prime },x_{2}^{\prime \prime },t^{\prime \prime
};x_{1}^{\prime },x_{2}^{\prime },t^{\prime }) &=&\stackunder{j=1}{\stackrel{%
2}{\prod }}\left[ m_{j}^{\prime \prime }m_{j}^{\prime }\right] ^{\frac{1}{4}%
}\exp \left[ \left. \frac{i}{2\hbar }\beta _{j}(t)\sqrt{m_{j}(t)}%
x_{j}^{2}\right| _{t^{\prime }}^{t^{\prime \prime }}\right]  \nonumber \\
&&\times K(Q_{1}^{\prime \prime },Q_{2}^{\prime \prime },t^{\prime \prime
};Q_{1}^{\prime },Q_{2}^{\prime },t^{\prime }),  \label{a.15}
\end{eqnarray}
where 
\begin{eqnarray}
K(Q_{1}^{\prime \prime },Q_{2}^{\prime \prime },t^{\prime \prime
};Q_{1}^{\prime },Q_{2}^{\prime },t^{\prime })\!\!\!\! &=&\!\!\!\!\int \!\!\!%
\stackunder{j=1}{\stackrel{2}{\prod }\!\!}\mathcal{D}Q_{j}\mathcal{D}%
P_{j}\exp \left\{ \!\!\frac{i}{\hbar }\int_{t^{\prime }}^{t^{\prime \prime
}}\!\!\left( P_{1}\stackrel{.}{\!Q}_{1}\!+\!P_{2}\stackrel{.}{\!Q}_{2}\!-\!%
\mathcal{H}(t)\right) dt\!\right\}  \nonumber \\
&&  \label{a.16}
\end{eqnarray}
is the propagator for the physical system governed by the Hamiltonian $%
\mathcal{H}(t)$ in the new conjugate variables.

If we make the following choices: 
\begin{equation}
\alpha (t)=\text{Const,}  \label{a.17a}
\end{equation}
and 
\begin{equation}
\beta _{j}(t)=-\frac{\stackrel{.}{m_{j}}(t)}{2\sqrt{m_{j}(t)}},\text{ \quad }%
j=1,2,  \label{a.17b}
\end{equation}
the terms in $P_{1}Q_{2}$ and $P_{2}Q_{1}$ are cancelled and the new
Hamiltonian (\ref{a.9}) takes the form: 
\begin{equation}
\mathcal{H}(t)=\stackunder{j=1}{\stackrel{2}{\sum }}\left[ \frac{P_{j}^{2}}{2%
}+\frac{1}{2}\Omega _{j}^{2}(t)Q_{j}^{2}-F_{j}(t)Q_{j}\right] +\Gamma
(t)Q_{1}Q_{2},  \label{a.18}
\end{equation}
where 
\begin{equation}
\left\{ 
\begin{array}{c}
\Omega _{1}(t)=\sqrt{\widetilde{\omega }_{1}^{2}(t)\cos ^{2}\alpha +%
\widetilde{\omega }_{2}^{2}(t)\sin ^{2}\alpha -\frac{\lambda (t)}{\sqrt{%
m_{1}(t)m_{2}(t)}}\sin (2\alpha )}, \\ 
\Omega _{2}(t)=\sqrt{\widetilde{\omega }_{1}^{2}(t)\sin ^{2}\alpha +%
\widetilde{\omega }_{2}^{2}(t)\cos ^{2}\alpha +\frac{\lambda (t)}{\sqrt{%
m_{1}(t)m_{2}(t)}}\sin (2\alpha )}, \\ 
F_{1}(t)=\sqrt{m_{1}(t)}f_{1}(t)\cos \alpha -\sqrt{m_{2}(t)}f_{2}(t)\sin
\alpha , \\ 
F_{2}(t)=\sqrt{m_{1}(t)}f_{1}(t)\sin \alpha +\sqrt{m_{2}(t)}f_{2}(t)\cos
\alpha , \\ 
\Gamma (t)=\frac{1}{2}\left[ \widetilde{\omega }_{1}^{2}(t)-\widetilde{%
\omega }_{2}^{2}(t)\right] \sin (2\alpha )+\frac{\lambda (t)}{\sqrt{%
m_{1}(t)m_{2}(t)}}\cos (2\alpha ),
\end{array}
\right.  \label{a.19}
\end{equation}
with 
\begin{equation}
\widetilde{\omega }_{j}^{2}(t)=\left[ \omega _{j}^{2}(t)+\frac{1}{4}\left( 
\frac{\stackrel{.}{m_{j}^{2}}(t)}{m_{j}^{2}(t)}-2\frac{\stackrel{..}{m_{j}}%
(t)}{m_{j}(t)}\right) \right] ;\,\quad j=1,2.  \label{a.20}
\end{equation}
Notice that, with the above canonical transformation, the coupling $\Gamma
(t)$ is a functional on the parameters of the original system.\ Then, the
separation of variables occurs for a wide but restricted class of two
coupled general driven time-dependent harmonic oscillators contrary to this
what has been asserted in Ref.\cite{Lo} .

It is clear that the separation of variables in Eq. (\ref{a.18}) requires
that $\Gamma (t)=0$, i.e. 
\begin{equation}
\lambda (t)=\frac{1}{2}\sqrt{m_{1}(t)m_{2}(t)}\left( \widetilde{\omega }%
_{2}^{2}(t)-\widetilde{\omega }_{1}^{2}(t)\right) \tan (2\alpha ).
\label{a.21}
\end{equation}

Then, by taking into account (\ref{a.21}), the propagator (\ref{a.16}) is
rewritten as: 
\begin{equation}
K(Q_{1}^{\prime \prime },Q_{2}^{\prime \prime },t^{\prime \prime
};Q_{1}^{\prime },Q_{2}^{\prime },t^{\prime })=\stackunder{j=1}{\stackrel{2}{%
\prod }}K(Q_{j}^{\prime \prime },t^{\prime \prime };Q_{j}^{\prime
},t^{\prime }),  \label{a.22}
\end{equation}
where 
\begin{equation}
K(Q_{j}^{\prime \prime },t^{\prime \prime };Q_{j}^{\prime },t^{\prime
})=\int \mathcal{D}Q_{j}\mathcal{D}P_{j}\exp \left\{ \frac{i}{\hbar }%
\int_{t^{\prime }}^{t^{\prime \prime }}\left( P_{j}\stackrel{.}{Q}%
_{j}-H_{j}(t)\right) dt\right\}  \label{a.23}
\end{equation}
is the propagator for a time-dependent harmonic oscillator with a
perturbative force described by the Hamiltonian 
\begin{equation}
H_{j}(t)=\frac{P_{j}^{2}}{2}+\frac{1}{2}\Omega
_{j}^{2}(t)Q_{j}^{2}-F_{j}(t)Q_{j}.  \label{a.24}
\end{equation}
Now, it is easy to write down the propagator (\ref{a.23}). Using the
generalized canonical transformations method, Chetouani and co-workers \cite
{Chetouani} have developped a very elegant way of calculating the
propagators of certain time-dependent one-dimensional physical systems. To
use this method, we apply the canonical tansformation, followed by a time
transformation, defined by 
\begin{equation}
Q_{j}=X_{j}\rho _{j}(t),\quad P_{j}=\mathcal{P}_{j}/\rho _{j}(t),\quad \frac{%
ds}{dt}=\rho _{j}^{-2}(t),  \label{a.25}
\end{equation}
where $\rho _{i}(t)$ is an arbitrary function without dimensions. After some
calculation, we find that 
\begin{equation}
K(Q_{j}^{\prime \prime },t^{\prime \prime };Q_{j}^{\prime },t^{\prime })\!=\!%
\frac{1}{\left( \rho _{j}^{\prime \prime }\rho _{j}^{\prime }\right) ^{1/2}}%
\exp \left\{ \!\frac{i}{2\hbar }\left( \frac{\stackrel{.}{\overline{\rho _{j}%
}}^{\prime \prime }}{\rho _{j}^{\prime \prime }}X_{j}^{\prime \prime 2}-%
\frac{\stackrel{.}{\overline{\rho _{j}}}^{\prime }}{\rho _{j}^{\prime }}%
X_{j}^{\prime 2}\right) \!\right\} K(X_{j}^{\prime \prime },s^{\prime \prime
};X_{j}^{\prime },s^{\prime }),  \label{a.26}
\end{equation}
where 
\begin{eqnarray}
K(X_{j}^{\prime \prime },s^{\prime \prime };X_{j}^{\prime },s^{\prime })
&=&\int \mathcal{D}Y_{j}\mathcal{DP}_{j}\exp \left\{ \frac{i}{\hbar }%
\int_{s^{\prime }}^{s^{\prime \prime }}\left[ \mathcal{P}_{j}\stackrel{.}{%
X_{j}}-\left( \frac{\mathcal{P}_{j}^{2}}{2}\right. \right. \right.  \nonumber
\\
&&\left. \left. \left. +\frac{1}{2}\left( \widetilde{\Omega }_{j}^{2}+%
\overline{\Omega }_{j}^{2}(s)\overline{\rho }_{j}^{4}\right) X_{j}^{2}-%
\overline{F}_{j}(s)\overline{\rho }_{j}^{3}X_{j}\right) ds\right] \right\} ,
\label{a.27}
\end{eqnarray}
with 
\begin{equation}
\widetilde{\Omega }_{j}^{2}=\left[ \frac{\stackrel{..}{\overline{\rho }}_{j}%
}{\overline{\rho }_{j}}-2\left( \frac{\stackrel{.}{\overline{\rho }}_{j}}{%
\overline{\rho }_{j}}\right) ^{2}\right] =\rho _{j}^{3}\stackrel{..}{\rho
_{j}}.  \label{a.28}
\end{equation}
In the above we have used the notation $\rho _{j}=\overline{\rho }%
_{j}(s)=\rho _{j}(t),$ $\stackrel{.}{\rho _{j}}=\frac{d\rho _{j}}{dt},$ $%
\stackrel{.}{\overline{\rho }}_{j}=\frac{d\overline{\rho }_{j}}{ds}$, $%
\overline{F}_{j}(s)=F_{j}(t)$ and $\Omega _{j}^{2}(t)=$.$\overline{\Omega }%
_{j}^{2}(s).$

A comparaison between the propagator (\ref{a.23}) and (\ref{a.26}) shows
that the space-time transformations (\ref{a.25}) has resulted in the
appearance of a phase and a quadratic term $\frac{1}{2}\widetilde{\Omega }%
_{j}^{2}X_{j}^{2}.\;$On the other hand the global time-dependent frequency
in Eq. (\ref{a.27}) depends on the adjustable parameter $\rho _{j}.$

If we now impose a constraint on $\rho _{j}$ by setting the global
time-dependent frequency appearing in Eq. (\ref{a.27}) equal to a constant: 
\begin{equation}
\widetilde{\Omega }_{j}^{2}+\overline{\Omega }_{j}^{2}(s)\overline{\rho }%
_{j}^{4}=\omega _{0j}^{2}=\text{Const},  \label{a.29}
\end{equation}
we recognize the propagator of Eq.(\ref{a.27}) as the expression given by
Feynman and Hibbs \cite{Feynman} for the forced harmonic oscillator with a
constant frequency and we therefore find for the propagator (\ref{a.26}) 
\begin{eqnarray}
K(Q_{j}^{\prime \prime },t^{\prime \prime };Q_{j}^{\prime },t^{\prime }) &=&%
\sqrt{\frac{\omega _{0j}}{2i\pi \hbar \rho _{j}^{\prime \prime }\rho
_{j}^{\prime }\sin \phi _{j}(t^{\prime \prime },t^{\prime })}}\exp \left\{ 
\frac{i}{2\hbar }\left( \frac{\stackrel{.}{\rho }_{j}^{\prime \prime }}{\rho
_{j}^{\prime \prime }}Q_{j}^{\prime \prime 2}-\frac{\stackrel{.}{\rho }%
_{j}^{\prime }}{\rho _{j}^{\prime }}Q_{j}^{\prime 2}\right) \right\} 
\nonumber \\
&&\times \exp \left\{ \frac{i\omega _{0j}}{2\hbar \sin \phi _{j}(t^{\prime
\prime },t^{\prime })}\left[ \left( \frac{Q_{j}^{\prime \prime 2}}{\rho
_{j}^{\prime \prime 2}}+\frac{Q_{j}^{\prime 2}}{\rho _{j}^{\prime 2}}\right)
\cos \phi _{j}(t^{\prime \prime },t^{\prime })\right. \right.  \nonumber \\
&&-\frac{2Q_{j}^{\prime \prime }Q_{j}^{\prime }}{\rho _{j}^{\prime \prime
}\rho _{j}^{\prime }}+\frac{2}{\omega _{0j}}\frac{Q_{j}^{\prime \prime }}{%
\rho _{j}^{\prime \prime }}\int_{t^{\prime }}^{t^{\prime \prime
}}G_{j}(t)\sin \phi _{j}(t,t^{\prime })dt  \nonumber \\
&&+\frac{2}{\omega _{0j}}\frac{Q_{j}^{\prime }}{\rho _{j}^{\prime }}%
\int_{t^{\prime }}^{t^{\prime \prime }}G_{j}(t)\sin \phi _{j}(t^{\prime
\prime },t)dt  \nonumber \\
&&\left. \left. -\frac{2}{\omega _{0j}^{2}}\int_{t^{\prime }}^{t^{\prime
\prime }}\!\int_{t^{\prime }}^{t}G_{j}(t)G_{j}(\tau )\sin \phi
_{j}(t^{\prime \prime },t)\sin \phi _{j}(\tau ,t^{\prime })d\tau dt\right]
\!\right\} ,  \nonumber \\
&&  \label{a.30}
\end{eqnarray}
where 
\begin{equation}
G_{j}(t)=F_{j}(t)\rho _{j}(t)\text{ \qquad and \qquad }\phi _{j}(t^{\prime
\prime },t^{\prime })=\omega _{0j}\int_{t^{\prime }}^{t^{\prime \prime }}%
\frac{dt}{\rho _{j}^{2}}.  \label{a.31}
\end{equation}

Substituting (\ref{a.28}) into (\ref{a.29}), we see that $\rho _j(t)$ is
solution to 
\begin{equation}
\stackrel{..}{\rho }_j+\Omega _j^2(t)\rho _j=\frac{\omega _{0j}^2}{\rho _j^3}%
,  \label{a.32}
\end{equation}
which is the well-known auxiliary equation \cite{Ermakov,Lewis} .

Inserting (\ref{a.30}) and (\ref{a.22}) in (\ref{a.15}) and using (\ref
{a.17b}), the final expression for the system of two time-coupled and driven
harmonic oscillators governed by the Hamiltonian (\ref{a.1}) is given by

\begin{eqnarray}
K(x_{1}^{\prime \prime },x_{2}^{\prime \prime },t^{\prime \prime
};x_{1}^{\prime },x_{2}^{\prime },t^{\prime })\!\!\! &=&\!\!\!\stackunder{j=1%
}{\stackrel{2}{\prod }}\sqrt{\frac{\left( m_{j}^{\prime \prime
}m_{j}^{\prime }\right) ^{\frac{1}{2}}\omega _{0j}}{2i\pi \hbar \rho
_{j}^{\prime \prime }\rho _{j}^{\prime }\sin \phi _{j}(t^{\prime \prime
},t^{\prime })}}\exp \left\{ -\frac{i}{4\hbar }\left. \frac{\stackrel{.}{m}%
_{j}(t)}{m_{j}(t)}x_{j}^{2}\right| _{t^{\prime }}^{t^{\prime \prime
}}\right\}  \nonumber \\
&&\!\!\!\!\!\!\!\!\!\!\!\!\times \exp \left\{ \frac{i}{2\hbar }\left( \frac{%
\stackrel{.}{\rho }_{j}^{\prime \prime }}{\rho _{j}^{\prime \prime }}%
Q_{j}^{\prime \prime 2}-\frac{\stackrel{.}{\rho }_{j}^{\prime }}{\rho
_{j}^{\prime }}Q_{j}^{\prime 2}\right) \right\}  \nonumber \\
&&\!\!\!\!\!\!\!\!\!\!\!\!\times \exp \left\{ \frac{i\omega _{0j}}{2\hbar
\sin \phi _{j}(t^{\prime \prime },t^{\prime })}\left[ \left( \frac{%
Q_{j}^{\prime \prime 2}}{\rho _{j}^{\prime \prime 2}}+\frac{Q_{j}^{\prime 2}%
}{\rho _{j}^{\prime 2}}\right) \cos \phi _{j}(t^{\prime \prime },t^{\prime
})\right. \right.  \nonumber \\
&&\!\!\!\!\!\!\!\!\!\!\!\!\!-\frac{2Q_{j}^{\prime \prime }Q_{j}^{\prime }}{%
\rho _{j}^{\prime \prime }\rho _{j}^{\prime }}+\frac{2}{\omega _{0j}}\frac{%
Q_{j}^{\prime \prime }}{\rho _{j}^{\prime \prime }}\int_{t^{\prime
}}^{t^{\prime \prime }}G_{j}(t)\sin \phi _{j}(t,t^{\prime })dt  \nonumber \\
&&\!\!\!\!\!\!\!\!\!\!\!\!\!+\frac{2}{\omega _{0j}}\frac{Q_{j}^{\prime }}{%
\rho _{j}^{\prime }}\int_{t^{\prime }}^{t^{\prime \prime }}G_{j}(t)\sin \phi
_{j}(t^{\prime \prime },t)dt  \nonumber \\
&&\!\!\!\!\!\!\!\!\!\!\!\!\!\!-\left. \left. \frac{2}{\omega _{0j}^{2}}%
\int_{t^{\prime }}^{t^{\prime \prime }}\int_{t^{\prime
}}^{t}G_{j}(t)G_{j}(\tau )\sin \phi _{j}(t^{\prime \prime },t)\sin \phi
_{j}(\tau ,t^{\prime })d\tau dt\right] \right\} ,  \nonumber \\
&&  \label{a.33}
\end{eqnarray}

Next, we turn our attention to the special case where $m_{j}(t)=m_{j}e^{%
\gamma t}$ and $\lambda (t)=rm_{1}m_{2}e^{\gamma t}.$ The propagator defined
by expression (\ref{a.15}) can be written 
\begin{equation}
K(x_{1}^{\prime \prime },x_{2}^{\prime \prime },t^{\prime \prime
};x_{1}^{\prime },x_{2}^{\prime },t^{\prime })=\stackunder{j=1}{\stackrel{2}{%
\prod }}\mathcal{K}(Q_{j}^{\prime \prime },t^{\prime \prime };Q_{j}^{\prime
},t^{\prime }),  \label{a.34}
\end{equation}
with 
\begin{eqnarray}
\mathcal{K}(Q_{j}^{\prime \prime },t^{\prime \prime };Q_{j}^{\prime
},t^{\prime }) &=&\sqrt{\frac{\omega _{0j}}{2i\pi \hbar \sigma _{j}^{\prime
\prime }\sigma _{j}^{\prime }\sin \phi _{j}(t^{\prime \prime },t^{\prime })}}%
\exp \left\{ \frac{i}{2\hbar }\left( \frac{\stackrel{.}{\sigma }_{j}^{\prime
\prime }}{\sigma _{j}^{\prime \prime }}Q_{j}^{\prime \prime 2}-\frac{%
\stackrel{.}{\sigma }_{j}^{\prime }}{\sigma _{j}^{\prime }}Q_{j}^{\prime
2}\right) \right\}  \nonumber \\
&&\!\!\!\ \times \exp \left\{ \frac{i\omega _{0j}}{2\hbar \sin \phi
_{j}(t^{\prime \prime },t^{\prime })}\left[ \left( \frac{Q_{j}^{\prime
\prime 2}}{m_{j}^{\prime \prime }\sigma _{j}^{\prime \prime 2}}+\frac{%
Q_{j}^{\prime 2}}{m_{j}^{\prime }\sigma _{j}^{\prime 2}}\right) \cos \phi
_{j}(t^{\prime \prime },t^{\prime })\right. \right.  \nonumber \\
&&\!\!\!\ -\frac{2Q_{j}^{\prime \prime }Q_{j}^{\prime }}{\sqrt{m_{j}^{\prime
\prime }m_{j}^{\prime }}\sigma _{j}^{\prime \prime }\sigma _{j}^{\prime }}+%
\frac{2}{\omega _{0j}}\frac{Q_{j}^{\prime \prime }}{\sqrt{m_{j}^{\prime
\prime }}\sigma _{j}^{\prime \prime }}\int_{t^{\prime }}^{t^{\prime \prime
}}G_{j}(t)\sin \phi _{j}(t,t^{\prime })dt  \nonumber \\
&&\ \!\!+\frac{2}{\omega _{0j}}\frac{Q_{j}^{\prime }}{\sqrt{m_{j}^{\prime }}%
\sigma _{j}^{\prime }}\int_{t^{\prime }}^{t^{\prime \prime }}G_{j}(t)\sin
\phi _{j}(t^{\prime \prime },t)dt  \nonumber \\
&&\!\!\!\ \left. \left. -\frac{2}{\omega _{0j}^{2}}\int_{t^{\prime
}}^{t^{\prime \prime }}\int_{t^{\prime }}^{t}G_{j}(t)G_{j}(\tau )\sin \phi
_{j}(t^{\prime \prime },t)\sin \phi _{j}(\tau ,t^{\prime })d\tau dt\right]
\right\} ,  \nonumber \\
&&  \label{a.35}
\end{eqnarray}
where $\sigma _{j}=\rho _{j}/\sqrt{m_{j}}.$

In the absence of the coupling, i. e. $\lambda (t)=0$ and furthermore $%
\alpha =0.$ Our system is then composed by two uncoupled harmonic
oscillators with time-dependent and different angular frequencies, masses
and forces. In this case, the propagator is

\begin{equation}
K(x_{1}^{\prime \prime },x_{2}^{\prime \prime },t^{\prime \prime
};x_{1}^{\prime },x_{2}^{\prime },t^{\prime })=\stackunder{j=1}{\stackrel{2}{%
\prod }}K(x_{j}^{\prime \prime },t^{\prime \prime };x_{j}^{\prime
},t^{\prime }),  \label{a.36}
\end{equation}
where 
\begin{eqnarray}
K(x_{j}^{\prime \prime },t^{\prime \prime };x_{j}^{\prime },t^{\prime
})\!\!\! &=&\!\!\!\sqrt{\frac{\omega _{0j}}{2i\pi \hbar \sigma _{j}^{\prime
\prime }\sigma _{j}^{\prime }\sin \phi _{j}(t^{\prime \prime },t^{\prime })}}%
\exp \left\{ \frac{i}{2\hbar }\left( m_{j}^{\prime \prime }\frac{\stackrel{.%
}{\sigma }_{j}^{\prime \prime }}{\sigma _{j}^{\prime \prime }}x_{j}^{\prime
\prime 2}-m_{j}^{\prime }\frac{\stackrel{.}{\sigma }_{j}^{\prime }}{\sigma
_{j}^{\prime }}x_{j}^{\prime 2}\right) \right\}  \nonumber \\
&&\times \exp \left\{ \frac{i\omega _{0j}}{2\hbar \sin \phi _{j}(t^{\prime
\prime },t^{\prime })}\left[ \left( \frac{x_{j}^{\prime \prime 2}}{\sigma
_{j}^{\prime \prime 2}}+\frac{x_{j}^{\prime 2}}{\sigma _{j}^{\prime 2}}%
\right) \cos \phi _{j}(t^{\prime \prime },t^{\prime })\right. \right. 
\nonumber \\
&&-\frac{2x_{j}^{\prime \prime }x_{j}^{\prime }}{\sigma _{j}^{\prime \prime
}\sigma _{j}^{\prime }}+\frac{2}{\omega _{0j}}\frac{x_{j}^{\prime \prime }}{%
\sigma _{j}^{\prime \prime }}\int_{t^{\prime }}^{t^{\prime \prime
}}G_{j}(t)\sin \phi _{j}(t,t^{\prime })dt  \nonumber \\
&&+\frac{2}{\omega _{0j}}\frac{x_{j}^{\prime }}{\sigma _{j}^{\prime }}%
\int_{t^{\prime }}^{t^{\prime \prime }}G_{j}(t)\sin \phi _{j}(t^{\prime
\prime },t)dt  \nonumber \\
&&-\left. \left. \frac{2}{\omega _{0j}^{2}}\int_{t^{\prime }}^{t^{\prime
\prime }}\!\int_{t^{\prime }}^{t}G_{j}(t)G_{j}(\tau )\sin \phi
_{j}(t^{\prime \prime },t)\sin \phi _{j}(\tau ,t^{\prime })d\tau dt\right]
\right\} ,  \nonumber \\
&&  \label{a.37}
\end{eqnarray}
is in agreement with the one obtained in Ref.\cite{Dhara} . This is clearly
proof that our exact results (\ref{a.30}) and (\ref{a.31}) are indeed the
correct quantum propagators for two time-dependent coupled and driven
harmonic oscillators.

In conclusion, this presentation contains two principal results. First, we
have succeeded in extending the path integral method to include a wide class
of a pair of time-dependent coupled and driven harmonic oscillators. By
explicit path integration, we were able to derive closed form expressions
for the Feynman propagator in a rigorous and explicit way.

Second, the correct propagators for the systems included in this study have
been obtained for the first time.

\end{document}